\documentclass[12pt,a4paper,aps,superscriptaddress,nofootinbib]{revtex4}
\usepackage[utf8]{inputenc}
\usepackage{graphicx}
\usepackage{amssymb}
\usepackage{textcomp}
\usepackage{amsmath}
\usepackage{tabularx}
\usepackage{bm}
\usepackage{times}
\usepackage{color}
\usepackage{ulem}
\usepackage{slashed}
\usepackage{multirow}
\usepackage{verbatim}
\usepackage{cancel}
\usepackage{subfigure}
\usepackage{epstopdf}
\usepackage{mathrsfs}
\usepackage{amsmath}

\usepackage[export]{adjustbox}
\def\ba{\begin{align}}
\def\ea{\end{align}}

\usepackage{tikz}
\usetikzlibrary{shapes,snakes}
\numberwithin{equation}{section}
\usepackage[colorlinks=true, pdfstartview=FitV, linkcolor=blue, citecolor=blue, urlcolor=blue]{hyperref}
\allowdisplaybreaks[4]

\def\lsim{\mathrel{\raise.3ex\hbox{$<$\kern-.75em\lower1ex\hbox{$\sim$}}}}
\def\gsim{\mathrel{\raise.3ex\hbox{$>$\kern-.75em\lower1ex\hbox{$\sim$}}}}

\definecolor{orange}{rgb}{1,0.5,0}

\begin{document}

\title{Quantum  Effective Dynamics of Papapetrou Spacetime}

\author{Xiao-Kan Guo}
\email{kankuohsiao@whu.edu.cn}
\affiliation{
School of Mathematics and Physics, Yancheng Institute of Technology, Jiangsu 224051, China
}
\author{Faqiang Yuan}
\email{yfq@mail.bnu.edu.cn (Corresponding Author)}
\affiliation{
School of Physics and Astronomy, Beijing Normal University,
Beijing 100875, China
}
\begin{abstract}
	In this paper, we investigate the quantum effective dynamics of  Papapetrou spacetime by using methods from loop quantum gravity.
	The Papapetrou spacetime is an extension of the Schwarzschild spacetime with an
	extra coupled anti-scalar field. By solving the equations of motion generated by the Hamiltonian constraint for Papapetrou spacetime, we can construct the quantum effective metric. The resulting effective metric for the quantum-corrected  Papapetrou spacetime demonstrates the quantum effects will give rise to a new wormhole throat, while the classical wormhole throat disappear in the case of extremely small mass.
\end{abstract}
\date{\today}

\maketitle
%\vspace{\stretch{3}}
%\thispagestyle{empty}
%\newpage
%\pagenumbering{arabic}
%%%%%%%%%%%%%%%%%%%%%%%%%%%%%%%%%%%%%%%%%%%%%%%%%%%%%%%%%
\section{Introduction}
Loop quantum gravity (LQG) is now one of the most popular approaches to quantise gravity.  This popularity is partially due to  the impressive mathematical rigour \cite{T}, as well as the clear conceptual foundations \cite{R}, of LQG. However, as a theory proposed  for physics, comparing its predictions with experiments or observations is the only way of testing its physical relevance. The developments of loop quantum cosmology (LQC) could provide possible tests of LQG; for example, the cosmological perturbation calculations in LQC can be compared with the cosmic microwave background observations \cite{AWW23}. On the other hand, the LQG-corrected effective black hole metrics have been recently derived using different methods \cite{BMM19,KSW20,ABV22,Z23,Z24}, which  inspires numerous recent works on the astrophysical properties of these LQG-corrected effective metrics (cf. the papers citing \cite{BMM19,KSW20,ABV22,Z23,Z24}).
In view of the recent advances in the ability of  astronomically observing both gravitational waves and  optical images of black holes, it is hoped that by comparing these calculations with future observations one could also find possible tests of LQG. 

To derive the effective metrics for quantum black holes, one can follow the seminal work by  Ashtekar, Olmedo, and Singh (AOS) \cite{AOS} where the quantum effective dynamics in a Schwarzschild spacetime is studied using the methods from LQC. By choosing a suitable quantisation scheme as in LQC, e.g. choosing the quantum parameters  used to regularise
the Hamiltonian constraint, one can write down the Hamiltonian equations of motion for the holonomy and flux variables of a black hole spacetime. If one could solve these equations of motion for the holonomies and fluxes, then the effective metrics could be readily constructed from these  variables. So far, the  studies on quantum-corrected effective metrics mainly focus on the spherically symmetric case, and different choices of quantisation schemes (cf. the Table 2 of \cite{Gan24} for all the existing quantisation schemes) lead to distinct quantum corrections to the classical Schwarzschild metric. 
One exception is the case of Janis-Newman-Winicour (JNW) spacetime \cite{JNW}, which is a spherically symmetric solution to the Einstein gravity minimally coupled with a scalar field and contains a naked singularity. The quantum effective dynamics of the JNW spacetime have been studied in \cite{ZZ20} and it is shown that the naked singularities can be avoided in the quantum effective dynamics.

In this paper, we consider the quantum effective dynamics as well as the effective metric for a case that is closely related to both the Schwarzschild and the JNW spacetimes, i.e. the Papapetrou spacetime \cite{Papa,MM18}. The papapetrou spacetime is a spherically symmetric solution to the Einstein gravity minimally coupled to anti-scalar fields (i.e. scalar fields whose energy-momentum tensor takes the opposite sign). The metric for the Papapetrou spacetime can be put into the form with exponential components, and such exponential metrics are found to represent traversable wormholes \cite{BNSV18,GQ24}. Therefore, the astrophysical properties of the Papapetrou spacetime differ from the black hole spacetime in several aspects. Importantly, the Papapetrou metric, together with the Schwarzschild and JNW metrics, can be treated a unified way with different limits of a single parameter in the metric components. In this work, we treat the Papapetrou metric as a limit of the JNW metric. We shall work in the Hamiltonian formalism and study the quantum effective dynamics of the Papapetrou spacetime using the same method as in  \cite{AOS,ZZ20}. It turns out that the Hamiltonian equations of motion for the Papapetrou spacetime are easier to solve than the JNW case, and the solutions of holonomy and flux variables give rise to a slightly simpler form of the effective metric for the quantum Papapetrou spacetime than that in the JNW case.

The effective Papapetrou metric obtained in this work represents a LQG-corrected  wormhole spacetime. Notice that the LQG corrections to the wormhole metrics have already been considered in the literature. For instance, in \cite{KLP10,KLP11} the polymer quantisation of the wormhole throat is performed in metric variables, but the dynamics is only solved numerically and the effective metric has not been derived. It is also found that the LQG holonomy corrections to the matter scource would make the wormhole throat disappear \cite{D11}.
More recently in \cite{Z24,BBSC25}, quantum-corrected wormhole solutions can be found as specific limits of the quantum-corrected Schwarzshild solutions obtained from the LQG  quantum effective dynamics, with which the present work shares many similarities but with the important difference of coupling to anti-scalar fields.
On the other hand, there are also works \cite{RGK23,Muniz1,Muniz2} studying the wormhole solutions to the Einstein equations modified by the LQC energy density derived in the effective Friedman equations. Although such a new source term can make the wormhole solutions traversable, this consideration seems problematic, since the LQC energy density is only derived from the quantised FRW model instead of the wormhole model. There is no guarantee that a direct quantisation of a wormhole spacetime could give rise to the same effective energy density as in LQC  even if it is assumed to exist in a quantised FRW universe. 
A less problematic approach is to include the LQG quantum corrections in the matter source, while retaining the classical geometric parts in the Einstein equations, cf. \cite{CNM24}.
The quantum effective metric for the exponential  metric obtained in this work clearly differs from all the wormhole metrics in these works.

There is also the problem of distinguishability between these quantum-corrected wormhole metrics in astronomical observations, so we should also study   the astrophyiscal properties of the quantum effective metric for Papapetrou spacetime. We plan to report these astrophysical studies in another work.

In Section \ref{II}, after a brief introduction to the classical Papapetrou spacetime, we obtain the classical Hamiltonian dynamics of the Papapetrou spacetime as a limit of the JNW classical dynamics. Then in Section \ref{III}, we study the quantum effective dynamics for the Papapetrou spacetime. By solving the quantum effective dynamics, we derive the quantum effective metric for the Papapetrou spacetime. In Section \ref{S4}, we show that the resulting quantum effective metric satisfy the wormhole conditions, and identify the wormhole throats. Section \ref{S5} concludes. 
%%%%%%%%%%%%%%%%%%%%%%%%%%%%%%%%%%%%%%%%%%%%%%%%%%%%%%%%%%%%
\section{Papapetrou spacetime}\label{II}
Let us start with the  metric line element for the JNW spacetime \cite{JNW},
\begin{equation}\label{1,1,}
	ds^2=-\Bigl(1-\frac{2M}{\nu r}\Bigr)^\nu dt^2+\Bigl(1-\frac{2M}{\nu r}\Bigr)^{-\nu} dr^2+\Bigl(1-\frac{2M}{\nu r}\Bigr)^{1-\nu} d\Omega^2,
\end{equation}
with a parameter $\nu=M/\sqrt{M^2+\sigma^2}$. Here, $\sigma$ is the scalar charge for the scalar field,
\begin{equation}
	\varphi=\sqrt{\frac{1-\nu^2}{2 \kappa}}\ln\Bigl(1-\frac{2M}{\nu r}\Bigr)
\end{equation}
with $\kappa=8 \pi G$, which is a solution to the massless Klein-Gordon equation on the background \eqref{1,1,}.
This metric \eqref{1,1,} is a spherically symmetric solution in the Einstein gravity minimally coupled with a scalar field.
In fact, when the parameter $\nu=1$, or equivalently $\sigma=0$, the metric \eqref{1,1,} reduces to the Schwarzschild black hole metric. An important feature of the JNW spacetime is an additional singularity at $r=2\sqrt{M^2+\sigma^2}$ for $\nu\neq1$, which represents a naked singularity. 

It is noticed in \cite{MM18} that in the limit $\nu\rightarrow\infty$, the JNW metric \eqref{1,1,} becomes the Papapetrou metric
\begin{equation}\label{3,3,}
	ds^2=-e^{-2M/r}dt^2+e^{2M/r}(dr^2+r^2d\Omega^2).
\end{equation}
In effect, this limit $\nu\rightarrow\infty$ can be equivalently taken by sending the scalar charge $\sigma$ to a purely imaginary value $iM$, i.e.  $\sigma\rightarrow iM$, by the definition of $\nu$. In this limit, we have $\varphi\rightarrow-i\sqrt{\frac{2}{\kappa}}M/r$. If we identify $\sqrt{\frac{2}{\kappa}} M/r\equiv{\varphi}$,
we then have the limit $\varphi\rightarrow i{\varphi}$, so that 
the energy-momentum tensor $T_{\mu\nu}(\varphi)$ of the scalar field, which is quadratic in $\varphi$, tends to $-T_{\mu\nu}({\varphi})$. Therefore, the Papapetrou metric is a solution to the Einstein gravity minimally coupled to the anti-scalar fields, i.e. the scalar fields with the opposite-sign energy momentum tensor. Compared to the JNW metric as well as the Schwarzschild metric,  the Papapetrou metric \eqref{3,3,} has a much simpler structure, since it has neither naked singularities nor event horizons.

\subsection{Classical Hamiltonian dynamics}
In this subsection, we use the method of radial evolution \cite{AOS} to study the classical Hamiltonian dynamics of  Papapetrou spacetime. 

Now that there is no interior region for  the Papapetrou spacetime, we should use the method of radial evolution for the exterior Schwarzschild spacetime \cite{AOS}. 
In this case, a spherically symmetric spacetime is foliated by time-like homogeneous surfaces whose isometry group is still $\mathbb{R} \times SO(3)$, and the time-like foliated surfaces have topology $\mathbb{R}\times S^2$. For convenience, we  work with a fiducial cell $\mathcal{C}\equiv(0,L_0)\times S^2$ to avoid divergences along the non-compact dimension.\footnote{Due to the homogeneity of the hypersurfaces, the physical information obtained from this elementary cell suffices to characterize the entire hypersurfaces. Since $L_0$ is arbitrarily chosen, the physical quantities derived from the theory should be independent of $L_0$. }
Since the intrinsic metric has signature $(-,+,+)$ on the time-like foliated surfaces,  the gauge group of internal rotations for the gravitational connections is now $\mathrm{SU(1,1)}$. The generators  of the Lie algebra  $\mathfrak{su}(1,1)$ are $\tilde{{\tau}}_1=i {\tau}_1,  \tilde{{\tau}}_2=i {\tau}_2,  \tilde{{\tau}}_3={\tau}_3$, where the ${\tau}_i$ are the generators of $\mathfrak{su}(2)$.
Then the gravitational Ashtekar connection and the conjugate densitized triad can be expressed as \cite{AOS}
\begin{align}
	\label{symmetry variables1}
	A_a^i \tilde{{\tau}_i} \mathrm{~d} x^a  =&\frac{\tilde{c}}{L_0} {\tau}_3 \mathrm{~d} x+i \tilde{b} {\tau}_2 \mathrm{~d} \theta- i \tilde{b} {\tau}_1 \sin \theta \mathrm{~d} \phi+{\tau}_3 \cos \theta \mathrm{~d} \varphi, \\
	E_i^a \tilde{{\tau}}^i \partial_a  =&\tilde{p}_c {\tau}_3 \sin \theta \partial_x+\frac{i \tilde{p}_b}{L_0} {\tau}_2 \sin \theta \partial_\theta-\frac{i \tilde{p}_b}{L_0} {\tau}_1 \partial_\varphi,
\end{align}
where $\tilde{b},\tilde{p}_b,\tilde{c},\tilde{p}_c$ are the canonically conjugate pairs for the symmetry reduced phase space satisfying 
\begin{equation}
	\{\tilde{b},\tilde{p}_b\}=-G\gamma,\quad \{\tilde{c},\tilde{p}_c\}=2G\gamma,
\end{equation}
with $\gamma$ being the Immirzi parameter. On the other hand, the scalar field and its conjugate momentum are reduced to the form
\begin{eqnarray}
	\label{reduced scalar}
	\varphi(z)=\tilde{\varphi}, \quad \pi_{\varphi}(z)= \frac{\tilde{\pi}_{\varphi}}{4\pi L_0} \sin\theta,
\end{eqnarray}
and the nonzero Possion braket between them is
\begin{eqnarray}
	\label{scalar Possion braket}
	\{\tilde{\varphi},\tilde{\pi}_\varphi\}=1.
\end{eqnarray}
By employing the above dynamical variables, we have the Hamiltonian constraint for the Einstein gravity minimally coupled with an anti-scalar field,
\begin{equation}
	\label{eHamiltonian}
	\tilde{H}(\tilde{N}) =-\frac{1}{2 G \gamma^2} \frac{\tilde{N} \operatorname{sgn}\left(\tilde{p}_b\right) i \tilde{b}}{\sqrt{\left|\tilde{p}_c\right|}}\left(\left(-\tilde{b}+\frac{\gamma^2}{\tilde{b}}\right) \tilde{p}_b+2 \tilde{c} \tilde{p}_c\right)+\frac{\tilde{N}}{2} \frac{\tilde{\pi}_{{\varphi}}^2 }{4 \pi i \left|\tilde{p}_b \right| \sqrt{\tilde{p}_c}}.
\end{equation} 
Now the strategy is to choose a particular form of the lapse $\tilde{N}_t$ associated with the time coordinate $t$, and then solve the corresponding Hamiltonian equations of motion of the dynamical variables, after which the  spacetime metric can be constructed by substituting these solutions into
\begin{equation}
	\label{metric1}
	\mathrm{d} s^2=-\tilde{N}_t^2 \mathrm{~d} t^2-\frac{\tilde{p}_b^2}{\left|\tilde{p}_c\right| L_0^2} \mathrm{~d} x^2+\left|\tilde{p}_c\right|\left(\mathrm{d} \theta^2+\sin ^2 \theta \mathrm{~d} \phi^2\right).
\end{equation} 
Note that the minus sign for the second term in \eqref{metric1} is due to the replacement $b\rightarrow i\tilde{b},p_b\rightarrow i\tilde{p}_b$ when compared to the interior Schwarzschild and JNW cases.

Let us choose the lapse as
\begin{equation}
	\tilde{N}_t=-i \kappa \gamma^2 |\tilde{p}_b| \sqrt{|\tilde{p}_c|},
\end{equation}
then the Hamiltonian constraint \eqref{eHamiltonian} becomes
\begin{equation}
	\label{eHamiltoniant}
	\tilde{H}(\tilde{N}_t) =-4 \pi \left[(-\tilde{b}^2+ \gamma^2)\tilde{p}_b^2 +2 \tilde{c} \tilde{p}_c \tilde{b}\tilde{p}_b\right] - G \gamma^2 \tilde{\pi}_\varphi^2.
\end{equation} 
Note that $\tilde{c}\tilde{p}_c$ and $\tilde{\pi}_\phi$ are constants of motion. Let us redefine the variables as $\tilde{c}\tilde{p}_c\equiv\gamma L_0 M$ and $\tilde{b}\tilde{p}_b\equiv\tilde{y}$, then the Hamiltonian equations of motion become
\begin{align}
	\frac{\mathrm{d} \tilde{y}}{\mathrm{d} t}=&\kappa \gamma^3 \tilde{p}_b^2,\label{ty}\\
	\frac{\mathrm{d} \tilde{p_b}}{\mathrm{d} t}=& -\kappa \gamma(-\tilde{b} \tilde{p}_b^2+ 2\gamma L_0 M\tilde{p_b}) ,\label{tpb}\\
	\frac{\mathrm{d} \tilde{c}}{\mathrm{d}t}=&-2 \kappa \gamma \tilde{c} \tilde{y},\label{tcequation}\\
	\frac{\mathrm{d} \tilde{p_c}}{\mathrm{d}t}=&2 \kappa \gamma \tilde{p_c} \tilde{y},\label{tpcequation}\\
	\frac{\mathrm{d} \tilde{\varphi}}{\mathrm{d}t}=&-\frac{\kappa \gamma^2}{4 \pi} \tilde{\pi}_\varphi.\label{tphiequation}
\end{align}
Since $\tilde{c}\tilde{p}_c$ is a constant of motion, we only need to solve one of the Eqs. \eqref{tcequation} and  \eqref{tpcequation}. According to this constant of motion, we can solve $\tilde{p_c}$ by
\begin{equation}
	\label{tpcc}
	\tilde{p}_c=\frac{\gamma L_0 M}{\tilde{c}}.
\end{equation}
On the other hand, since the Hamiltonian constraint is preserved during the evolution, we only need to solve one of Eqs. \eqref{ty} and \eqref{tpb}. According to the constraint $\tilde{H}(\tilde{N}_t)=0$, $\tilde{p}_b^2$ can be expressed as
\begin{equation}
	\label{tpb2}
	\tilde{p}_b^2=\frac{1}{\gamma^2}(\tilde{y}+y_+)(\tilde{y}+y_-),
\end{equation}
where $y_{\pm}=-\gamma L_0 M \pm \gamma L_0 \sqrt{M^2+{G \tilde{\pi}_\varphi^2}/{4 \pi L_0^2}}$. Substituting \eqref{tpb2} to \eqref{ty}, we have 
\begin{eqnarray}
	\label{ty2}
	\frac{\mathrm{d} \tilde{y}}{\mathrm{d} t}=\kappa \gamma (\tilde{y}+y_+)(\tilde{y}+y_-).
\end{eqnarray}
Thus, the equations of motion that we need to solve are  \eqref{tcequation}, \eqref{tphiequation}, \eqref{tpcc}, \eqref{tpb2}, and \eqref{ty2}.

To recover the Papapetrou spacetime, we need to take the following  limit for the scalar charge
\begin{equation}\sigma:=\frac{\sqrt{\frac{G}{4 \pi}}\tilde{\pi}_\varphi}{L_0} \longrightarrow i M.\end{equation} 
In this limit, the equations of motion that we need to solve are read as
\begin{align}
	\frac{\mathrm{d} \tilde{y}}{\mathrm{d} t}=&\kappa \gamma (\tilde{y}-\gamma L_0 M)^2,\label{ty2p}\\
	\tilde{p}_b=&\frac{1}{\gamma}(\gamma L_0 M-\tilde{y}),\label{tpb2p}\\
	\frac{\mathrm{d} \tilde{\varphi}}{\mathrm{d} t}=&-i \sqrt{2\kappa} \gamma^2 L_0 M, \label{tphip}
\end{align}
together with Eqs. \eqref{tcequation} and \eqref{tpcc}.

The solution of \eqref{ty2p} is simply 
\begin{equation}
	\label{tpb2ps}
	\tilde{y}(t)=\gamma L_0 M-\frac{1}{\kappa \gamma t}.
\end{equation}
Substituting this into \eqref{tpb2p}, we get
\begin{equation}
	\label{tpb2pbs}
	\tilde{p_b}(t)=\frac{1}{\kappa \gamma^2 t}.
\end{equation}
According to the definition of $\tilde{y}$, $\tilde{b}$ is then 
\begin{equation}
	\label{tb2bs}
	\tilde{b}(t)=\kappa \gamma^3 L_0 M t - \gamma.
\end{equation}
Next, considering the equation of motion for $\tilde{c}$, we have
\begin{equation}
	\label{tcequation1}
	-\frac{1}{2}\frac{\mathrm{d} (\ln \tilde{c})}{\mathrm{d}t}=\kappa \gamma \tilde{y}.
\end{equation}
By denoting $T\equiv-\frac{1}{2} \ln \tilde{c}$, we have  the solution to \eqref{tcequation1},
\begin{equation}
	\label{tcsolution}
	\tilde{c}(t)= e^{-2 T(t)},
\end{equation}
where 
\begin{equation}
	\label{Tsolution}
	T(t)=\kappa \gamma^2 L_0 M t- \ln t-\frac{1}{2} \ln (\kappa^2 \gamma^5 L_0^3 M).
\end{equation}
By \eqref{tpcc}, we have also  the solution of $\tilde{p}_c(t)$,
\begin{equation}
	\label{tpcsolutionp}
	\tilde{p}_c(t)=\gamma L_0 M e^{2 T(t)}=\frac{1}{\kappa^2 \gamma^4 L_0^2} \frac{1}{t^2} e^{2 \kappa \gamma^2 L_0 M t}.
\end{equation}
Finally, the solution to \eqref{tphip} is simply
\begin{eqnarray}
	\label{tphips}
	\tilde{\varphi}(t)=-i\sqrt{2 \kappa} \gamma L_0 M t.
\end{eqnarray}

We have now  the solutions of all the dynamical variables expressed in terms of the time 
parameter $t$. By substituting them into \eqref{metric1} we can  reconstruct the spacetime metric. 
In order to compare with the standard form of the Papapetrou solution, we make a coordinate transformation, $\tilde{\tau}=\frac{1}{\kappa \gamma^2 L_0 t}$; substituting it into Eqs. \eqref{tpb2pbs}, \eqref{tpcsolutionp}, \eqref{tphips} and the lapse $\tilde{N}_t$, we have  
\begin{equation}
	\tilde{p}_b^2(\tilde{\tau})=L_0^2 \tilde{\tau}^2,\quad\tilde{p}_c(\tilde{\tau})=\tilde{\tau}^2 e^{\frac{2M}{\tilde{\tau}}},\quad\tilde{\varphi}(\tilde{\tau})= -i \sqrt{\frac{2}{\kappa}} \frac{M}{\tilde{\tau}},\quad\tilde{N}_t^2 =-\kappa^2 \gamma^4 L_0^2 \tilde{\tau}^4 e^{\frac{2M}{\tilde{\tau}}}.\label{dytau}
\end{equation}
With these solutions \eqref{dytau} in hand, we see that the metric  \eqref{metric1} becomes 
\begin{align}
	d s^2=&\kappa^2 \gamma^4 L_0^2 \tilde{\tau}^4 e^{\frac{2M}{\tilde{\tau}}} \mathrm{~d} t^2- e^{-\frac{2M}{\tilde{\tau}}} \mathrm{~d} x^2+\tilde{\tau}^2 e^{\frac{2M}{\tilde{\tau}}}\left(\mathrm{d} \theta^2+\sin ^2 \theta \mathrm{~d} \phi^2\right)= \nonumber\\
	=&\kappa^2 \gamma^4 L_0^2 \tilde{\tau}^4 e^{\frac{2M}{\tilde{\tau}}}  \frac{\mathrm{d^2 t}}{\mathrm{~d} \tilde{\tau}^2}\mathrm{~d} \tilde{\tau}^2- e^{-\frac{2M}{\tilde{\tau}}} \mathrm{~d} x^2+\tilde{\tau}^2 e^{\frac{2M}{\tilde{\tau}}}\left(\mathrm{d} \theta^2+\sin ^2 \theta \mathrm{~d} \phi^2\right)=\nonumber\\
	=&e^{\frac{2M}{\tilde{\tau}}} \mathrm{d}\tilde{\tau}^2- e^{-\frac{2M}{\tilde{\tau}}} \mathrm{~d} x^2+\tilde{\tau}^2 e^{\frac{2M}{\tilde{\tau}}}\left(\mathrm{d} \theta^2+\sin ^2 \theta \mathrm{~d} \phi^2\right)\label{metric2},
\end{align}
By replacing $x \rightarrow t, \tilde{\tau} \rightarrow r$, we obtain the standard form of the Papapetrou metric \eqref{3,3,}. We therefore see that the classical Hamiltonian dynamics obtained from the radial evolution method is consistent for the Papapetrou spacetime.
%%%%%%%%%%%%%%%%%%%%%%%
\section {Quantum effective  dynamics} \label{III}
In this section, we turn to the quantum effective dynamics of  Papapetrou spacetime. The strategy is to use the polymerisations of connection variables to include the quantum effects like in LQC, and then solve the resulting effective dynamics. This method has been shown to be successful for the Schwarzschild and JNW spacetimes \cite{AOS,ZZ20}. 

We first consider the holonomy along the $x$-direction with length $\tilde{\delta}_c L_0$
\begin{equation}
	\label{c holo}
	h_{x}^{\tilde{\delta}_c}=e^{\int_0^{\tilde{\delta}_c L_0} \frac{\tilde{c}}{L_0} \tilde{\tau}_3 \mathrm{d} \theta}=\cos(\frac{\tilde{\delta}_c \tilde{c}}{2})\mathbb{I} +2 \tilde{\tau}_3 \sin(\frac{ \tilde{\delta}_c \tilde{c}}{2}).
\end{equation}
To extract a scalar quantity from the holonomies, we make the following replacing
\begin{equation}
	\label{replacec}
	\tilde{c} \rightarrow -\frac{\mathrm{Tr}\left[h_{x}^{2\tilde{\delta}_c}\tilde{\tau}_3\right] }{ \tilde{\delta}_c}=\frac{\sin(\tilde{\delta}_c \tilde{c})}{\tilde{\delta}_c}.
\end{equation}
Then consider the holonomies along the equator and longitude with length $\tilde{\delta_b}$,
\begin{align}
	h_{\theta}^{\tilde{\delta}_b}&=e^{\int_0^{\tilde{\delta}_b} i \tilde{b} \tilde{\tau}_2 \mathrm{d} \theta}=\cos(\frac{i \tilde{\delta}_b \tilde{b}}{2})\mathbb{I} +2 \tilde{\tau}_2 \sin(\frac{i \tilde{\delta}_b \tilde{b}}{2}),\\
	h_{\phi}^{\tilde{\delta}_b}&=e^{\int_0^{\tilde{\delta}_b} (-i \tilde{b} \tilde{\tau}_1) \mathrm{d} \phi}=\cos(\frac{i \tilde{\delta}_b \tilde{b}}{2})\mathbb{I} -2 \tilde{\tau}_1 \sin(\frac{i \tilde{\delta}_b \tilde{b}}{2}).\label{holonomies}
\end{align}     
Similarly, we make the following replacing
\begin{equation}
	\label{replaceb}
	\tilde{b} \rightarrow -\frac{\mathrm{Tr}\left[h_{\theta}^{2\tilde{\delta}_b}\tilde{\tau}_2\right] }{i \tilde{\delta}_b}=\frac{\sin(i\tilde{\delta}_b \tilde{b})}{i \tilde{\delta}_b}=\frac{\sinh(\tilde{\delta}_b \tilde{b})}{\tilde{\delta}_b}.
\end{equation}
Note that now the replacing in \eqref{replaceb} uses a $\sinh$ function rather than a $\sin$ function. The quantum
parameters $\tilde{\delta}_b$ and $\tilde{\delta}_c$ are constants or functions on the phase space. The distinct ways in which they depend on the dynamical variables will result in different dynamics. In our analysis, we require them to be constants on the phase space. Substituting these replacements into the Hamiltonian constraint, we have the effective Hamiltonian 
\begin{equation}
	\label{eHamiltoniant1}
	\tilde{H}_{\mathrm{eff}}(\tilde{N}_t) =-4 \pi \left[2 \gamma L_0 M \tilde{x} -\tilde{x}^2 + \gamma^2\tilde{p}_b^2 \right] - G \gamma^2 \tilde{\pi}_\varphi^2,
\end{equation} 
where we have defined $\tilde{x}\equiv\frac{\sinh(\tilde{\delta}_b \tilde{b})}{\tilde{\delta}_b} \tilde{p}_b$, $\gamma L_0 M \equiv \frac{\sin(\tilde{\delta}_c \tilde{c})}{\tilde{\delta}_c} \tilde{p}_c$, and the lapse is chosen as $\tilde{N}_t=-i \kappa \gamma^2 |\tilde{p}_b| \sqrt{|\tilde{p}_c|}$.

From the effective Hamiltonian \eqref{eHamiltoniant1}, we obtain the  equation of motion for $\tilde{x}$,
\begin{equation}
	\label{ex}
	\frac{\mathrm{d} \tilde{x}}{\mathrm{d}t}=\kappa \gamma^3 \cosh(\tilde{\delta}_b \tilde{b}) \tilde{p}_b^2.
\end{equation}
By employing the constraint $\tilde{H}_{\mathrm{eff}}(\tilde{N}_t)=0$, we can solve $\tilde{p}_b^2$ as
\begin{equation}
	\label{ttpb2}
	\tilde{p}_b^2=\frac{1}{\gamma^2} (\tilde{x}^2- 2\gamma L_0 M \tilde{x} -\tilde{c}),
\end{equation}
where $\tilde{c}=({G \gamma^2 \tilde{\pi}_\varphi^2})/{4\pi}$. Substituting \eqref{ttpb2} into \eqref{ex} and  taking the limit $\sigma \rightarrow i M$, we obtain
\begin{equation}
	\label{ex2}
	\frac{\mathrm{d} \tilde{x}}{\mathrm{d}t}=\kappa \gamma ^3 \sqrt{1+\sinh^2(\tilde{\delta}_b \tilde{b})} \tilde{p}_b^2 
	= \kappa \gamma (\gamma L_0 M-\tilde{x}) \sqrt{\tilde{d}\tilde{x}^2- 2\gamma L_0 M \tilde{x} +\gamma^2 L_0^2 M^2},
\end{equation}
where $\tilde{d}= 1+\gamma^2 \tilde{\delta}_b^2$. The solution to \eqref{ex2} is 
\begin{equation}
	\label{xts}
	\tilde{x}(t)=\gamma L_0 M-\frac{\gamma L_0 M \sqrt{\tilde{d}-1}}{\sqrt{\tilde{d}-1}+\tanh(\sqrt{\tilde{d}-1}(\kappa \gamma^2 L_0 M t-1))}.
\end{equation}
By using \eqref{ttpb2}, we also have the solution of $\tilde{p}_b$,
\begin{equation}
	\label{ttpbp}
	\tilde{p}_b(t)=\frac{L_0 M \sqrt{\tilde{d}-1}}{\sqrt{\tilde{d}-1}+\tanh(\sqrt{\tilde{d}-1}(\kappa \gamma^2 L_0 M t-1))}.
\end{equation}
Then, by the definition of $\tilde{x}$, we get the solution of $\tilde{b}$,
\begin{equation}
	\label{ttbp}
	\frac{\sinh(\tilde{\delta_b} \tilde{b})}{\tilde{\delta_b}}=\frac{\tilde{x}}{{\tilde{p}_b}}=\frac{ \gamma \tanh(\sqrt{\tilde{d}-1}(\kappa \gamma^2 L_0 M t-1))}{\sqrt{\tilde{d}-1}}.
\end{equation}
In the above formulas, we have obviously $\sqrt{\tilde{d}-1}=\gamma\tilde{\delta}_b$, which will be used in the final expression.

Next, consider the equation of motion for $\tilde{c}$, that is,
\begin{equation}
	\label{ttcequation}
	\frac{\mathrm{d} \tilde{c}}{\mathrm{d}t}=-2\kappa\gamma \frac{\sin(\tilde{\delta}_c \tilde{c})}{\tilde{\delta}_c} \tilde{x}.
\end{equation}
This is equivalent to
\begin{equation}
	\label{cequation2}
	\frac{\mathrm{d}}{\mathrm{d}t}{\left[-\frac{1}{2} \ln(\tan \frac{\tilde{\delta}_c \tilde{c}}{2})\right]}=\kappa\gamma \tilde{x}.
\end{equation}
By defining $\tilde{T}\equiv-\frac{1}{2} \ln (\tan \frac{\tilde{\delta}_c\tilde{c}}{2})+\frac{1}{2} \ln(\frac{ \tilde{\delta}_c}{2})$, we can formally solve \eqref{cequation2} to get
\begin{equation}
	\label{ttc}
	\tan \left(\frac{\tilde{\delta}_c \tilde{c}(\tilde{T})}{2}\right)  =\frac{ \tilde{\delta}_c}{2} e^{-2 \tilde{T}}.
\end{equation}
By the definition of $\gamma L_0 M$, the solution of $\tilde{p}_c$ is then
\begin{equation}
	\label{ttpc}
	\tilde{p}_c(\tilde{T})  =  \gamma L_0 M \left(e^{2 \tilde{T}}+\frac{ \tilde{\delta}_c^2}{4} e^{-2 \tilde{T}}\right),
\end{equation}
where $\tilde{T}(t)$ satisfies the differential equation
\begin{equation}
	\label{tTe}
	\frac{\mathrm{d}\tilde{T}}{\mathrm{d}t}=\kappa \gamma \tilde{x}(t)=\kappa \gamma \left(\gamma L_0 M-\frac{\gamma L_0 M \sqrt{\tilde{d}-1}}{\sqrt{\tilde{d}-1}+\tanh(\sqrt{\tilde{d}-1}(\kappa \gamma^2 L_0 M t-1))}\right).
\end{equation}
We can solve \eqref{tTe} for
\begin{align}
	\tilde{T}(t)=&{-\frac{1}{2-\tilde{d}} \ln\left[\cosh\left(\sqrt{\tilde{d}-1}(\kappa \gamma^2 L_0 M t-1)\right)+ \frac{1}{\sqrt{\tilde{d}-1}} \sinh\left(\sqrt{\tilde{d}-1}(\kappa \gamma^2 L_0 M t-1)\right) \right]} +\nonumber\\
	&+\frac{\kappa \gamma^2 L_0 M t}{2-\tilde{d}}+ \frac{1}{2} \ln(\frac{M}{\gamma L_0}).\label{Tts}
\end{align}
The solution of $\tilde{\varphi}$ is the same as the classical solution. It is easy to see that the effective solutions will go back to the classical solutions when $\tilde{\delta_b} \rightarrow 0$ and $\tilde{\delta_c} \rightarrow 0$. 

Up to this point, we have obtained the solutions of the quantum effective dynamics for  Papapetrou spacetime. Substituting these solutions into \eqref{metric1}, we  can get the effective metric, which has a very long expression. In order to facilitate further calculations, we implement a suitable coordinate transformation
\begin{eqnarray}
	\label{tprime}
	{\tilde{\tau}} =\frac{M}{1+\frac{\tanh(\sqrt{\tilde{d}-1}(\kappa \gamma^2 L_0 M t-1))}{\sqrt{\tilde{d}-1}}}.
\end{eqnarray}
In terms of this new coordinate ${\tilde{\tau}}$, the dynamical variables $\tilde{p}_b$, $\tilde{p}_c$, $\varphi$ and $\frac{\mathrm{d} t}{\mathrm{d}\tilde{\tau}}$ are  respectively given by
\begin{align}
	\tilde{p}_b(\tilde{\tau})&=L_0 \tilde{\tau} ,\label{54}\\
	\tilde{p}_c(\tilde{\tau})&=M^2\left[ \frac{\tilde{\tau}^2(1+\gamma \tilde{\delta}_b(\frac{M}{\tilde{\tau}}-1))^{1+\frac{1}{\gamma \tilde{\delta}_b}} (1-\gamma \tilde{\delta}_b(\frac{M}{\tilde{\tau}}-1))^{1-\frac{1}{\gamma \tilde{\delta}_b}}e^2}{M^2}\right]^{\frac{1}{1-\gamma^2 \tilde{\delta}_b^2}} +\nonumber\\
	&+\frac{\gamma^2 L_0^2 \tilde{\delta}_c^2}{4}\left[ \frac{M^2}{\tilde{\tau}^2(1+\gamma \tilde{\delta}_b(\frac{M}{\tilde{\tau}}-1))^{1+\frac{1}{\gamma \tilde{\delta}_b}} (1-\gamma \tilde{\delta}_b(\frac{M}{\tilde{\tau}}-1))^{1-\frac{1}{\gamma \tilde{\delta}_b}}e^2}\right]^{\frac{1}{1-\gamma^2 \tilde{\delta}_b^2}}, \label{55}\\
	\tilde{\varphi}&= \frac{-i\sqrt{\frac{2 }{\kappa}}}{\gamma}\left[\frac{1}{2\gamma \tilde{\delta}_b} \ln(1+\gamma \tilde{\delta}_b (\frac{M}{\tilde{\tau}}-1))-\frac{1}{2\gamma \tilde{\delta}_b}\ln(1-\gamma \tilde{\delta}_b (\frac{M}{\tilde{\tau}}-1))+1\right], \label{56}\\
	\frac{\mathrm{d} t}{\mathrm{d}\tilde{\tau}}&=-\frac{1}{\kappa \gamma^2 L_0 \tilde{\tau}^2} \frac{1}{1-\gamma^2 \tilde{\delta}_b^2(\frac{M}{\tilde{\tau}}-1)^2}.\label{57}
\end{align} 
The quantum-corrected effective spacetime metric can be constructed from
\begin{equation}
	\label{metric2}
	\mathrm{d} s^2=-g_{xx} \mathrm{~d} x^2+g_{\tilde{\tau} \tilde{\tau}} \mathrm{~d} \tilde{\tau}^2+ g_{\Omega \Omega}\left(\mathrm{d} \theta^2+\sin ^2 \theta \mathrm{~d} \phi^2\right)
\end{equation}     
with 
\begin{align}
	\label{conp}
	g_{xx}&=\frac{\tilde{p}_b^2(\tilde{\tau})}{|\tilde{p}_c(\tilde{\tau})| L_0^2},\\
	g_{\tilde{\tau} \tilde{\tau}}&= \kappa^2 \gamma^4 \tilde{p}_b^2(\tilde{\tau}) |\tilde{p}_c(\tilde{\tau})| (\frac{\mathrm{d} t}{\mathrm{d}\tilde{\tau}})^2,\\
	g_{\Omega \Omega}&=|\tilde{p}_c(\tilde{\tau})|.
\end{align}
By replacing $x \rightarrow t, \tilde{\tau} \rightarrow r$, we obtain the resulting effective Papapetrou metric. From \eqref{54},\eqref{55},\eqref{56},\eqref{57}, we can verify that the effective metric will go back to the classical Papapetrou metric \eqref{3,3,} when $\tilde{\delta_b} \rightarrow 0$ and $\tilde{\delta_c} \rightarrow 0$. To see this more clearly, we can expand the metric components of \eqref{metric2} into a Taylor series of $\tilde{\delta}_b$  and $\tilde{\delta}_c$, and check the quantum corrections order by order. In the Appendix \ref{AnnA}, we present the  Taylor expansion of the effective metric around $\tilde{\delta}_b=0=\tilde{\delta}_c$ (i.e. the MacLaurin expansion)  to the first order.

%in order to ensure that physically meaningful quantities, such as the spacetime metric, are independent of the choice of the fiducial cell size, we need to examine how the two quantum parameters $\tilde{\delta_b},\tilde{\delta_c}$ depend on $L_0$. Notice that  the connection component $\tilde{b}$ is invariant, but $\tilde{c}$ changes via $\tilde{c} \rightarrow \alpha \tilde{c}$, under the rescaling of $L_0 \rightarrow \alpha L_0$ \cite{AOS}. Since $\tilde{b}$ and $\tilde{c}$ enter the effective equations through trigonometric functions of $\tilde{\delta_b} b$ and hyperbolic functions of $\tilde{\delta_c} c$ respectively, ensuring the cell independence of their solutions requires specifying $\tilde{\delta_b} b$ and $\tilde{\delta_c} c$ in such a way that they do not depend on the rescaling of fiducial cell. Since $\tilde{\delta_b}$ is the component of the connection in the angular direction, $\tilde{\delta_b}$ represents an angular, so we can choose $\tilde{\delta_b}=2\pi$. Since $\tilde{\delta_c} c$ is the component of the connection in the $x$ direction, $\tilde{\delta_c} L_0$ represents a length in the $x$ direction, so we choose $\tilde{\delta_c}  L_0=\sqrt{\Delta}$, where $\Delta$ is the minimum non-zero eigenvalue of the area operator in LQG. It is easy to see that this choice satisfies the requirement of 'fiducial cell independence'. 
%%%%%%%%%%%%%%%%%%%%%%%%
\section{Wormhole conditions}\label{S4}
Since the classical Papapetrou spacetime can be interpreted as a wormhole spacetime, we expect that the quantum-corrected Papapetrou metric still represent a wormhole with quantum corrections. To this end, let us look at the wormhole conditions \cite{BNSV18,GQ24,MT88} for the effective Papapetrou metric.

Consider the area of the spherical surfaces of constant $\tilde{\tau}$ coordinate; in \eqref{metric2} this is simply
\begin{equation}
	\label{A}
	A(\tilde{\tau})=4\pi\tilde{p}_c(\tilde{\tau}).
\end{equation} 
We have to show that this area $A(\tilde{\tau})$ is a concave function of $\tilde{\tau}$.

In order to calculate its derivatives with respective to $\tilde{\tau}$, we find  the form \eqref{ttpc} is more convenient, so we have
\begin{align}
	\frac{\mathrm{d} A(\tilde{\tau})}{\mathrm{d} \tilde{\tau}}=&8\pi \gamma L_0 M \left(e^{2 \tilde{T}}- \frac{\tilde{\delta_c}^2}{4}e^{-2 \tilde{T}} \right) \frac{\mathrm{d} \tilde{T}}{\mathrm{d} t} \frac{\mathrm{d} t}{\mathrm{d} \tilde{\tau}}=8\pi \kappa \gamma^2 L_0 M \left(e^{2 \tilde{T}}- \frac{\tilde{\delta_c}^2}{4}e^{-2 \tilde{T}} \right) \tilde{x} \frac{\mathrm{d} t}{\mathrm{d} \tilde{\tau}}= \nonumber\\
	=&8\pi \kappa \gamma^2 L_0 M \left(e^{2 \tilde{T}}- \frac{\tilde{\delta_c}^2}{4}e^{-2 \tilde{T}} \right) \frac{\gamma L_0 M \tanh(\sqrt{\tilde{d}-1}(\kappa \gamma^2 L_0 M t-1))}{\sqrt{\tilde{d}-1}+\tanh(\sqrt{\tilde{d}-1}(\kappa \gamma^2 L_0 M t-1))}\times \nonumber\\ &\times \frac{-1}{\kappa \gamma^2 L_0 \tilde{\tau}^2} \frac{1}{1-\gamma^2 \tilde{\delta}_b^2(\frac{M}{\tilde{\tau}}-1)^2}.
\end{align}
It is easy to see that $\frac{\mathrm{d} A(\tilde{\tau})}{\mathrm{d} \tilde{\tau}}=0$ at two places:
\begin{align}
	\label{dA1}
	e^{2 \tilde{T}}- \frac{\tilde{\delta_c}^2}{4}e^{-2 \tilde{T}}=0,\quad&\Longrightarrow\quad T_0=\frac{1}{2} \ln(\frac{\tilde{\delta_c}}{2}),\\
	\label{dA2}
	\tanh(\sqrt{\tilde{d}-1}(\kappa \gamma^2 L_0 M t-1))=0,\quad&\Longrightarrow\quad t_0= \frac{1}{\kappa \gamma^2 L_0 M}.
\end{align} 
Then we calculate the second derivative of $A(\tilde{\tau})$ with respective to $\tilde{\tau}$ at the above two points,  $\tilde{\tau}(\tilde{T} = T_0)$ and $\tilde{\tau}(t = t_0)$. Here the functional dependence of $\tilde{\tau}(\tilde{T})$ is determined by \eqref{Tts} and \eqref{tprime} and the functional relation $\tilde{\tau}(t)$ is given by \eqref{tprime}. The solution $\tilde{\tau}(\tilde{T}=T_0)$ is a consequence of the quantum correction associated with the quantum parameter $\tilde{\delta}_c$, and we have
\begin{align}
	\left. \frac{\mathrm{d}^2 A(\tilde{\tau})}{\mathrm{d} \tilde{\tau}^2} \right|_{\tilde{\tau}(\tilde{T}=T_0)}=&\left.16\pi \gamma L_0 M \left(e^{2 \tilde{T}}+ \frac{\tilde{\delta_c}^2}{4}e^{-2 \tilde{T}} \right) (\frac{\mathrm{d} \tilde{T}}{\mathrm{d} t} \frac{\mathrm{d} t}{\mathrm{d} \tilde{\tau}})^2 \right|_{\tilde{\tau}(\tilde{T}=T_0)} \nonumber\\ & + \left.8\pi \kappa \gamma^2 L_0 M \left(e^{2 \tilde{T}}- \frac{\tilde{\delta_c}^2}{4}e^{-2 \tilde{T}} \right)  \frac{\mathrm{d}^2 \tilde{T}}{\mathrm{d} \tilde{\tau}^2}\right|_{\tilde{\tau}(\tilde{T}=T_0)} \nonumber\\
	=&16\pi \gamma L_0 M \left(e^{2 \tilde{T}}+ \frac{\tilde{\delta_c}^2}{4}e^{-2 \tilde{T}} \right)  (\frac{\gamma L_0 M \tanh(\sqrt{\tilde{d}-1}(\kappa \gamma^2 L_0 M t-1))}{\sqrt{\tilde{d}-1}+\tanh(\sqrt{\tilde{d}-1}(\kappa \gamma^2 L_0 M t-1))})^2 \times\nonumber\\ &\times \left. (\frac{-1}{\kappa \gamma^2 L_0 \tilde{\tau}^2} \frac{1}{1-\gamma^2 \tilde{\delta}_b^2(\frac{M}{\tilde{\tau}}-1)^2})^2 \right|_{\tilde{\tau}(\tilde{T}=T_0)} >0,
\end{align}
Therefore, $\tilde{\tau}(\tilde{T}=T_0)$ satisfies the ``flare-out'' condition. From \eqref{tprime}, it follows that another solution $\tilde{\tau}(t = t_0)$ corresponds to $\tilde{\tau} = M$, which confirms that, in the effective spacetime, the point $\tilde{\tau} = M$ corresponds to the classical wormhole throat \cite{NS}. 
Likewise, we have
\begin{align}
	\left. \frac{\mathrm{d}^2 A(\tilde{\tau})}{\mathrm{d} \tilde{\tau}^2} \right|_{\tilde{\tau} = M}&=\left.8\pi \gamma L_0 M \frac{\mathrm{d} \tilde{x}}{\mathrm{d} t} \left(e^{2 \tilde{T}}- \frac{\tilde{\delta_c}^2}{4}e^{-2 \tilde{T}} \right) (\frac{\mathrm{d} t}{\mathrm{d} \tilde{\tau}})^2 \right|_{\tilde{\tau} = M} + \left. \tilde{x} \frac{ \mathrm{d} \left(8\pi \kappa \gamma^2 L_0 M  \left(e^{2 \tilde{T}}- \frac{\tilde{\delta_c}^2}{4}e^{-2 \tilde{T}} \right)  \frac{\mathrm{d}^2 \tilde{t}}{\mathrm{d} \tilde{\tau}} \right)}{\mathrm{d} \tilde{\tau}}\right|_{\tilde{\tau} = M} \nonumber\\
	&=\left. 8 \pi \kappa \gamma^4 L_0 M  \cosh(\tilde{\delta_b} \tilde{b}) \tilde{p}_b^2 \left(e^{2 \tilde{T}}- \frac{\tilde{\delta_c}^2}{4}e^{-2 \tilde{T}} \right)  (\frac{-1}{\kappa \gamma^2 L_0 \tilde{\tau}^2} \frac{1}{1-\gamma^2 \tilde{\delta}_b^2(\frac{M}{\tilde{\tau}}-1)^2})^2 \right|_{\tilde{\tau} = M} \nonumber\\ 
	&=\left. \frac{8\pi L_0}{\kappa M} \left(e^{2 \tilde{T}}- \frac{\tilde{\delta_c}^2}{4}e^{-2 \tilde{T}} \right) \right|_{\tilde{\tau} = M} \nonumber\\
	&=e^{\frac{2}{1-\gamma^2 \tilde{\delta}_b^2}} \frac{M}{\gamma L_0} -\frac{\tilde{\delta_c}^2}{4} e^{\frac{-2}{1-\gamma^2 \tilde{\delta}_b^2}} \frac{\gamma L_0}{M} .\label{6767}
\end{align}
Obviously, this second derivatives \eqref{6767} is not positive definite. For the case of $M> \frac{\gamma \tilde{\delta_c} L_0}{2} e^{\frac{-2}{1-\gamma^2 \tilde{\delta}_b^2}}$, we see that $\left. \frac{\mathrm{d}^2 A(\tilde{\tau})}{\mathrm{d} \tilde{\tau}^2} \right|_{\tilde{\tau} = M}>0$. However, for the case of $M \leq \frac{\gamma \tilde{\delta_c} L_0}{2} e^{\frac{-2}{1-\gamma^2 \tilde{\delta}_b^2}}$, the second derivatives $\left. \frac{\mathrm{d}^2 A(\tilde{\tau})}{\mathrm{d} \tilde{\tau}^2} \right|_{\tilde{\tau} = M} \leq 0$ means the classical wormhole throat disappear.  

Furthermore, it is easy to see that all metric components are finite at $\tilde{\tau}(\tilde{T} = T_0)$ or $\tilde{\tau} = M$, and the diagonal components are non-zero. Therefore, the quantum effective dynamics shows the classical wormhole throat disappear in the case of extremely small mass. On the other hand, quantum effects will give rise to a new wormhole throat at $\tilde{\tau}(\tilde{T} = T_0)$.  

Notice that in the above calculations we have not specified the numerical values of the quantum parameters $\tilde{\delta}_b$ and $\tilde{\delta}_c$ which definitely affect the final conclusions.
 We can choose these parameters according to their relations to the minimum area
gap in LQG.  
Recall that the area operator in LQG has a smallest nonzero eigenvalue given by $\Delta = 2 \sqrt{3} \pi \gamma G \hbar$ \cite{T}.  Since we have worked with a fiducial cell of size $L_0$ to avoid divergence in the non-compact direction of the foliated surface (of topology $\mathbb{R}\times S^2$), for the $x-\theta$ and $x-\phi$ faces of the fiducial cell, the area operator acts on the eigenstates of the holonomies $h_{\theta}^{\tilde{\delta}_b}$ and $h_{\phi}^{\tilde{\delta}_b}$, giving eigenvalues  $\tilde{\delta}_b \pi \gamma G \hbar$. By requiring this area to  be equal to the minimal eigenvalue $\Delta$, we obtain $\tilde{\delta}_b = 2\sqrt{3}$~\cite{AB}. On the other hand, since $\tilde{\delta}_c L_0$ represents a length in the $x$ direction, we choose $\tilde{\delta}_c L_0 = \sqrt{\Delta}$~\cite{LR}. It is easy to verify that this choice satisfies the requirement of fiducial cell independence.

\begin{figure}
		\centering
		\includegraphics[height=8cm,width=12cm]{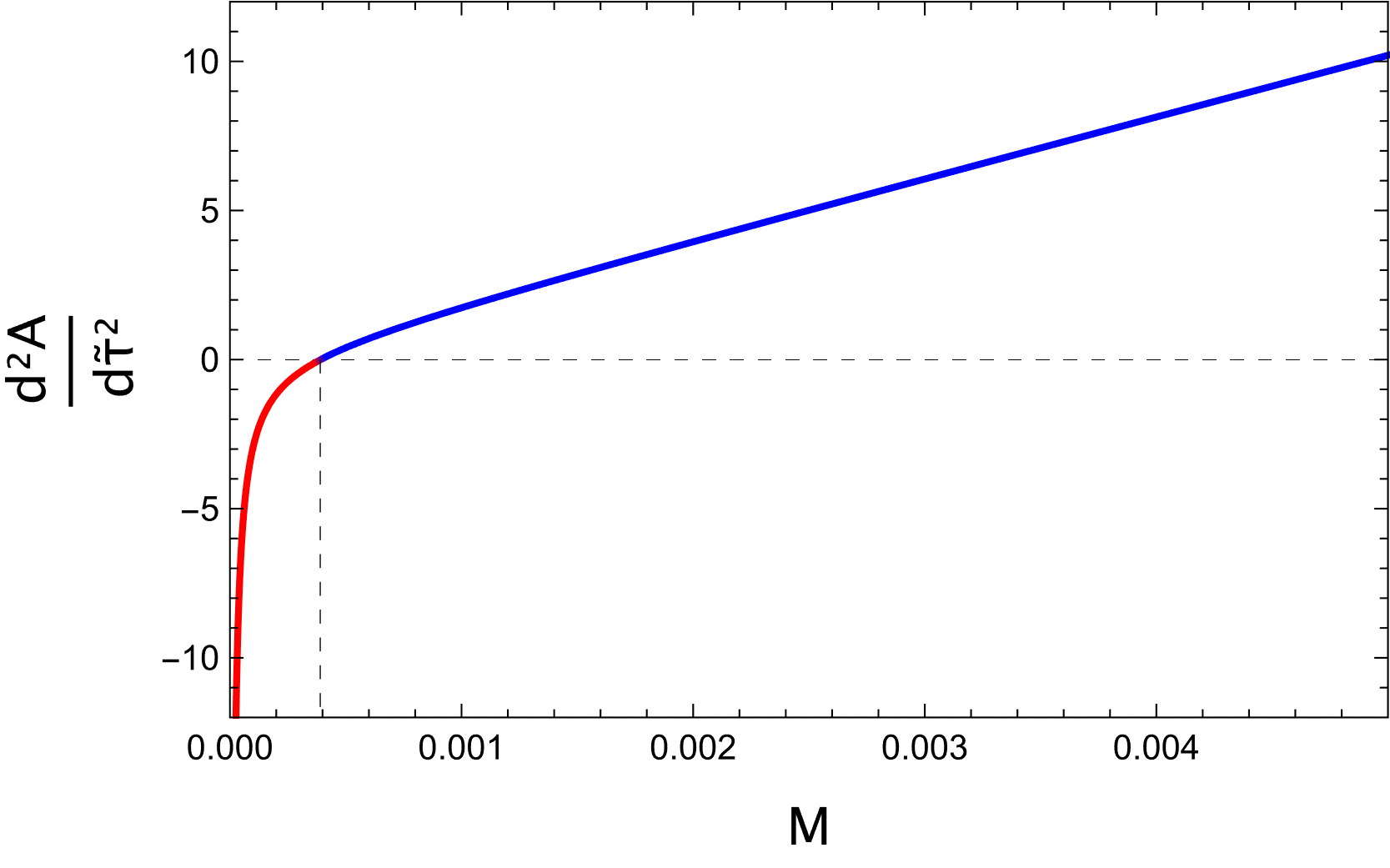}
		\caption{Plot of $\left. \frac{\mathrm{d}^2M A(\tilde{\tau})}{\mathrm{d} \tilde{\tau}^2} \right|_{\tilde{\tau}=M}$ as a function of the mass $M$. The parameters in this plot are chosen as $\gamma=0.2375$, $L_0=1$, $\tilde{\delta_b}=2 \sqrt{3}$, $\tilde{\delta_c}  L_0=(2 \sqrt{3} \pi \gamma G \hbar)^{1/2}$, $\hbar=1$ and $G=1$.}\label{F1}
	\end{figure}

With this choice of quantum parameters, let us check our results by specific numerical results. To this end, we further specify the  Immirzi parameter  to be $\gamma = 0.2375$ (through the study of black hole entropy).
 The Figure \ref{F1} shows the second derivative of the spherical surfaces area $A(\tilde{\tau})$ at $\tilde{\tau} = M$ as a function of the mass $M$ in the quantum-corrected spacetime. As illustrated in the figure, the``flare-out" condition associated with the classical wormhole throat is no longer satisfied when $M \leq 0.00039 \sqrt{G \hbar}$.
	It is important to emphasize that the condition $\tilde{\delta}_b$ and $\tilde{\delta}_c L_0$ being independent of the fiducial cell rescaling ensures that our result does not depend on the choice of $L_0$. Given that $\tilde{\delta}_c L_0$ is typically of the order of the Planck length—as in our setting where $\tilde{\delta}_c L_0 = (2 \sqrt{3} \pi \gamma G \hbar)^{1/2}$—we conclude that the classical wormhole throat will disappear for extremely small masses. 
	
Another important issue is the travesability of the effective wormhole, which requires a necessary condition that the null energy condition must be violated so as to stabilize the wormhole throat \cite{MT88}.
 Since we have only considered the wormhole throat without specifying the matter content at the worm hole throat, we cannot directly check the energy conditions in this effective spacetime. Nevertheless, we can, as in \cite{CNM24}, suppose that the geometric part in the Einstein equation remains intact and the quantum corrections appear in the matter part of Einstein equation. This  way, we can effectively study the energy-momentum tensor for matter by simply calculating the Einstein tensor.\footnote{ For the classical Papapetrou spacetime, one can use this strategy to obtain the energy-momentum tensor by the computing the Einstein tensor, and the result is that the null energy condition (and other energy conditions) is violated for the  Papapetrou spacetime, cf. \cite{NS}.}
 
 Notice that it is difficult to derive the inverse function $t(\tilde{T})$ of Eq.~\eqref{Tts} analytically.
For simplicity, we only compute the explicit forms of the non-zero components of the Einstein tensor at  $\tilde{\tau} = M$,
	\begin{align}
		\label{Gxx}
		G^x_x=&\frac{4 e^{\frac{2}{1 - \gamma^2 \tilde{\delta}_ b^2}} \left(4 e^{\frac{4}{1 - \gamma^2 \tilde{\delta}_ b^2}} M^2 - 3 L_0^2 \gamma^2 \tilde{\delta}_ c^2\right)}{\left(4 e^{\frac{4}{1 - \gamma^2 \tilde{\delta}_ b^2}} M^2 + L_0^2 \gamma^2 \tilde{\delta}_ c^2\right)^2}	,\\
		G^{\tilde{\tau}}_{\tilde{\tau}}=&-G^{\theta}_{\theta}=-G^{\phi}_{\phi}=-\frac{4 e^{\frac{2}{1 - \gamma^2 \tilde{\delta}_ b^2}}}{4 e^{\frac{4}{1 - \gamma^2 \tilde{\delta}_ b^2}} M^2 + L_0^2 \gamma^2 \tilde{\delta}_ c^2} .
	\end{align}
    It leads to the energy density and pressures,
	\begin{align}
		\label{rho}
		\rho=&-G^x_x=-\kappa \frac{4 e^{\frac{2}{1 - \gamma^2 \tilde{\delta}_ b^2}} \left(4 e^{\frac{4}{1 - \gamma^2 \tilde{\delta}_ b^2}} M^2 - 3 L_0^2 \gamma^2 \tilde{\delta}_ c^2\right)}{\left(4 e^{\frac{4}{1 - \gamma^2 \tilde{\delta}_ b^2}} M^2 + L_0^2 \gamma^2 \tilde{\delta}_ c^2\right)^2},\\
		p_r=&-p_t=-\kappa \frac{4 e^{\frac{2}{1 - \gamma^2 \tilde{\delta}_ b^2}}}{4 e^{\frac{4}{1 - \gamma^2 \tilde{\delta}_ b^2}} M^2 + L_0^2 \gamma^2 \tilde{\delta}_ c^2}
	\end{align} 		
	It is evident that in the limit $\tilde{\delta}_b \rightarrow 0$ and $\tilde{\delta}_c \rightarrow 0$, the energy density, radial pressure, and tangential pressure recover their classical values at the wormhole throat
	\begin{align}
		\label{rpc}
		\rho = p_r=-p_t=-\frac{\kappa}{e^2 M^2}
	\end{align} 
which is consistent with the classical results \cite{NS}.
    Then, we have
    \begin{align}
    	\label{een}
    	\rho+p_r&=-\kappa \frac{8 e^{\frac{2}{1 - \gamma^2 \tilde{\delta}_ b^2}} \left(4 e^{\frac{4}{1 - \gamma^2 \tilde{\delta}_ b^2}} M^2 - L_0^2 \gamma^2 \tilde{\delta}_ c^2\right)}{\left(4 e^{\frac{4}{1 - \gamma^2 \tilde{\delta}_ b^2}} M^2 + L_0^2 \gamma^2 \tilde{\delta}_ c^2\right)^2}	, \\ 
    	\rho+p_t&=\rho+ p_r+2p_t=\kappa \frac{16 e^{\frac{4}{1 - \gamma^2 \tilde{\delta}_ b^2}} L_0^2 \gamma^2 \tilde{\delta}_ c^2}{\left(4 e^{\frac{4}{1 - \gamma^2 \tilde{\delta}_ b^2}} M^2 + L_0^2 \gamma^2 \tilde{\delta}_ c^2\right)^2} >0.
    \end{align}
 %Recall that the energy conditions are given as, \\
    %Null Energy Condition(NEC): $\rho+p_r \geqslant 0, \rho+p_t \geqslant 0$ \\
    %Weak Energy Condition(WEC): $\rho \geqslant 0, \rho+p_r \geqslant 0, \rho+p_t \geqslant 0$ \\
    %Strong Energy Condition(SEC): $\rho \geqslant 0, \rho+p_t \geqslant 0, \rho+p_r+2 p_t \geqslant 0$  \\
    It is straightforward to verify that when $M > \frac{\gamma \tilde{\delta}_c L_0}{2} e^{\frac{-2}{1 - \gamma^2 \tilde{\delta}_b^2}}$, we have $\rho + p_r < 0$ at the throat $\tilde{\tau} = M$. As a result, the null energy condition, $\rho+p_r \geqslant 0, \rho+p_t \geqslant 0$, is violated at this point. Therefore, the quantum-corrected Papapetrou spacetime still admits a traversable wormhole throat supported by exotic matter at  $\tilde{\tau} = M$. In contrast, when $M \leq \frac{\gamma \tilde{\delta}_c L_0}{2} e^{\frac{-2}{1 - \gamma^2 \tilde{\delta}_b^2}}$,  all energy conditions are satisfied at the point $\tilde{\tau} = M$. As a result, the exotic matter vanishes and the classical wormhole throat also disappears, which is consistent with the results of flare-out conditions and also with  the result given in \cite{D11}. 

\section{Conclusion and discussion}\label{S5}
We have studied in this work the quantum effective dynamics of  Papapetrou spacetime. The Papapetrou spacetime  is treated as a limit of the JNW spacetime, whereby we obtain its classical Hamiltonian dynamics  as a limiting case of the Hamiltonian dynamics for the JNW spacetime. Then we solve its quantum effective dynamics and obtain the quantum-corrected  metric for  Papapetrou spacetime. Although the final expression of the  quantum-corrected effective metric is  a little complicated, the metric components only involve elementary functions, so that it is apt for numerics. We see that the resulting quantum effective Papapetrou metric represents a wormhole spacetime. In the small mass limit, quantum effect will make the classical wormhole throat disappear, which is accompanied by the restoration of null energy condition.

There remain many problems for further investigations. Here, we mention some aspects:

We did not calculate the travesability of the new wormhole throat, due to the difficulty of analytically solving the inverse function. Further numerical works on this issue are required. Of course, more reliable results on the travesability should be based on the study of concrete matter fields in this spacetime, which is a challenging topic for future works.

We have worked with two quantum numbers  $\tilde{\delta}_b,\tilde{\delta}_c$, with a specific choice of forms,
but there exist various other choices with particular features (cf. the Table 2 of \cite{Gan24} for a summary). Since there is no consensus on such a choice, we could also consider the quantum effective dynamics for Papapetrou spacetime with other choices of quantum parameters. Another important extension is to study the covariant quantum effective dynamics for Papapetrou spacetime as in \cite{Z24}.

We should study the astrophysical properties of the quantum corrected Papapetrou metric, following the lines of \cite{MM18,BNSV18,TTAS22,MRC23,MMSJ24}. At present, there is no guarantee that future astronomical observations would exactly confirm the astrophyical properties of the quantum-corrected black holes, so theoretical calculations about the astrophysical properties of alternative quantum-corrected spacetimes are equally important.

%\begin{enumerate}
%\item Different choices of the lapse......
%\item Rotational......
%\item Charged, covariant......
%\item Astrophysical properties......
%\end{enumerate}

%%%%%%%%%%%%%%%%%%%%%%%%%%%%%%%%%%%%%%%%%%%%%%%%%%%%%%%%%%%%%%
\section*{Acknowledgements} 
%The author thanks XYZ for helpful comments.
X.-K. Guo is supported by the funding for school-level research projects (xjr2024030)
of Yancheng Institute of Technology. Faqiang Yuan is supported
by the National Natural Science Foundation of China
(Grant Nos.12275022).
%%%%%%%%%%%%%%%%%%%%%%%%%%%%%%%%%%%%%%
\appendix
\section{Effective metric with first-order correction }\label{AnnA}
In this Appendix, we present the results of the Taylor expansion of the metric components of \eqref{metric2}. These expressions show the first-order quantum corrections to the classical Papapetrou metric.

First, we expand the metric components $\tilde{\delta}_b$ around $\tilde{\delta}_b=0$,
\begin{align}
	g_{\Omega \Omega}=&\frac{\mathrm{e}^{-\frac{2M}{\tilde{\tau}}} \tilde{\delta}_c^2 L_0^2 M^2 \gamma^2}{4 \tilde{\tau}^2} + \mathrm{e}^{\frac{2M}{\tilde{\tau}}} \tilde{\tau}^2 + 
	\frac{1}{12 \tilde{\tau}^5}\mathrm{e}^{-\frac{2M}{\tilde{\tau}}} \gamma^2 \left(-\tilde{\delta}_c^2 L_0^2 M^2 \gamma^2 + 4 \mathrm{e}^{\frac{4M}{\tilde{\tau}}} \tilde{\tau}^4\right) \times\nonumber\\
	&\times\left(2M^3 - 9M^2 \tilde{\tau} + 18M \tilde{\tau}^2 -5\tilde{\tau}^3 -6\tilde{\tau}^3 \ln\frac{M}{\tilde{\tau}}\right)  \tilde{\delta}_b^2 + o(\tilde{\delta}_b^2),\label{A1}\\
	g_{xx}=&\frac{1}{
		\mathrm{e}^{\frac{2M}{\tilde{\tau}}} + \frac{\mathrm{e}^{-\frac{2M}{\tilde{\tau}}} \tilde{\delta}_c^2 L_0^2 M^2 \gamma^2}{4 \tilde{\tau}^4}
	}- \nonumber\\
	&- \frac{4 \left(
		\mathrm{e}^{\frac{2M}{\tilde{\tau}}} \gamma^2 \tilde{\tau} \left(
		-\tilde{\delta}_c^2 L_0^2 M^2 \gamma^2 + 4 \mathrm{e}^{\frac{4M}{\tilde{\tau}}} \tilde{\tau}^4
		\right)  \left(2M^3 - 9M^2 \tilde{\tau} + 18M \tilde{\tau}^2 -5\tilde{\tau}^3 -6\tilde{\tau}^3 \ln\frac{M}{\tilde{\tau}}\right) \tilde{\delta}_b^2
		\right)}{
		3 \left(
		\tilde{\delta}_c^2 L_0^2 M^2 \gamma^2 + 4 \mathrm{e}^{\frac{4M}{\tilde{\tau}}} \tilde{\tau}^4
		\right)^2
	} + o(\tilde{\delta}_b^2),\label{A2}\\
	g_{\tilde{\tau}\tilde{\tau}}=&\mathrm{e}^{\frac{2M}{\tilde{\tau}}} + \frac{\mathrm{e}^{-\frac{2M}{\tilde{\tau}}} \tilde{\delta}_c^2 L_0^2 M^2 \gamma^2}{4 \tilde{\tau}^4}+\nonumber \\
	&+\frac{1}{12 \tilde{\tau}^7} \mathrm{e}^{-\frac{2M}{\tilde{\tau}}}  4 \mathrm{e}^{\frac{4M}{\tilde{\tau}}} \gamma^2 \tilde{\tau}^4 \left(
	2 M^3 - 3 M^2 \tilde{\tau} + 6 M \tilde{\tau}^2 + \tilde{\tau}^3 -6\tilde{\tau}^3 \ln\frac{M}{\tilde{\tau}}
	\right) \tilde{\delta}_b^2 -\nonumber\\
	&-\frac{1}{12 \tilde{\tau}^7} \mathrm{e}^{-\frac{2M}{\tilde{\tau}}} \tilde{\delta}_c^2 L_0^2 M^2 \gamma^4 \left(
	2 M^3 -15 M^2 \tilde{\tau} +30 M \tilde{\tau}^2 -  11 \tilde{\tau}^3 -6\tilde{\tau}^3 \ln\frac{M}{\tilde{\tau}}
	\right) \tilde{\delta}_b^2 + o(\tilde{\delta}_b^2).\label{A3}
\end{align}
Then, we perform the Taylor expansion of \eqref{A1}\eqref{A2}\eqref{A3} around $\tilde{\delta}_c=0$ to get 
\begin{align}
	g_{\Omega \Omega}=&\mathrm{e}^{\frac{2M}{\tilde{\tau}}} \tilde{\tau}^2+\frac{\mathrm{e}^{\frac{2M}{\tilde{\tau}}} \left(
		\gamma^2 \tilde{\delta}_b^2 \left(
		2M^3 - 9M^2 \tilde{\tau} + 18M \tilde{\tau}^2 -5\tilde{\tau}^3 -6\tilde{\tau}^3 \ln\frac{M}{\tilde{\tau}}
		\right)
		\right)}{3 \tilde{\tau}} +\nonumber\\
	&+ \frac{\mathrm{e}^{-\frac{2M}{\tilde{\tau}}}
		L_0^2 M^2 \gamma^2  \tilde{\delta}_c^2}{4 \tilde{\tau}^2}+ o(\tilde{\delta}_b^2)+o(\tilde{\delta}_c^2),\quad\label{A7}\\
	g_{xx}=&\mathrm{e}^{-\frac{2M}{\tilde{\tau}}}+\frac{1}{3} \mathrm{e}^{-\frac{2M}{\tilde{\tau}}} \left(
	\frac{
		\gamma^2 \tilde{\delta}_b^2 \left(
		2M^3 - 9M^2 \tilde{\tau} + 18M \tilde{\tau}^2 -5\tilde{\tau}^3 -6\tilde{\tau}^3 \ln\frac{M}{\tilde{\tau}}
		\right)
	}{\tilde{\tau}^3}
	\right) -\nonumber\\
	&- \frac{
		\mathrm{e}^{-\frac{6M}{\tilde{\tau}}}
		L_0^2 M^2 \gamma^2  \tilde{\delta}_c^2
	}{4 \tilde{\tau}^4} + o(\tilde{\delta}_b^2)+o(\tilde{\delta}_c^2),\label{A8}\\
	g_{\tilde{\tau}\tilde{\tau}}=&\mathrm{e}^{\frac{2M}{\tilde{\tau}}}+\frac{1}{3} \mathrm{e}^{\frac{2M}{\tilde{\tau}}} \left( \frac{
		\gamma^2 \tilde{\delta}_b^2 \left(
		2 M^3 - 3 M^2 \tilde{\tau} + 6 M \tilde{\tau}^2 + \tilde{\tau}^3 -6\tilde{\tau}^3 \ln\frac{M}{\tilde{\tau}}
		\right)
	}{\tilde{\tau}^3}
	\right) +\nonumber\\
	&+ \frac{\mathrm{e}^{-\frac{2M}{\tilde{\tau}}} \tilde{\delta}_c^2 L_0^2 M^2 \gamma^2}{4 \tilde{\tau}^4}+ o(\tilde{\delta}_b^2)+o(\tilde{\delta}_c^2).\label{A9}
\end{align}
Finally, using the replacement $x \rightarrow t, \tilde{\tau} \rightarrow r$, we obtain the Papapetrou metric with first-order corrections. Although the quantum correction terms are a little complicated, these terms contain only elementary functions.

%%%%%%%%%%%%%%%%%%%%%%%%%%%%%%%%%%%%%%%%%%%%%%%%%%%%%%%%%%%
\bibliographystyle{amsalpha}

\end{document}